\documentclass[12pt, a4paper]{article}
\usepackage[english]{babel}
\usepackage{amsmath,amssymb}
\usepackage{cite,mcite}
\usepackage{array}
\usepackage{xspace}
\usepackage{endnotes}
\usepackage{epsfig}
\usepackage{rotate}
\begin{document}
\newcommand{\eVdist}{\kern-0.045em}
\newcommand{\ev}{\ensuremath{\text{e}\eVdist\text{V}}}
\newcommand{\kev}{\ensuremath{\text{ke}\eVdist\text{V}}}
\newcommand{\mev}{\ensuremath{\text{Me}\eVdist\text{V}}}
\newcommand{\gev}{\ensuremath{\text{Ge}\eVdist\text{V}}}
\newcommand{\tev}{\ensuremath{\text{Te}\eVdist\text{V}}}
\let\eV=\ev
\let\keV=\kev
\let\MeV=\mev
\let\GeV=\gev
\let\TeV=\tev
\newcommand{\bev}{\ensuremath{\mathbf{e}\eVdist\mathbf{V}}}
\newcommand{\bkev}{\ensuremath{\mathbf{ke}\eVdist\mathbf{V}}}
\newcommand{\bmev}{\ensuremath{\mathbf{Me}\eVdist\mathbf{V}}}
\newcommand{\bgev}{\ensuremath{\mathbf{Ge}\eVdist\mathbf{V}}}
\newcommand{\btev}{\ensuremath{\mathbf{Te}\eVdist\mathbf{V}}}
\def\smallfrac#1#2{\hbox{${{#1}\over {#2}}$}}
\def\as{\alpha_s}
\def\bea{\begin{eqnarray}}
\def\eea{\end{eqnarray}}
\def\blackbox{\vrule height7pt width5pt depth2pt}
\def\matele#1#2#3{\langle {#1} \vert {#2} \vert {#3} \rangle }
\def\VertL{\Vert_{\Lambda}}\def\VertR{\Vert_{\Lambda_R}}
\def\Real{\Re e}\def\Imag{\Im m}
\def\SZP{\hbox{S0$'$}}\def\DZP{\hbox{D0$'$}}\def\DMP{\hbox{D-$'$}}
\def\MS{\hbox{$\overline{\rm MS}$}}
\def\ms{\hbox{$\overline{\scriptstyle\rm MS}$}}
\def\half{\hbox{${1\over 2}$}}\def\third{\hbox{${1\over 3}$}}
\def\QMS{Q$_0$\MS}
\def\QDIS{Q$_0$DIS}
\catcode`@=11 
\def\toinf#1{\mathrel{\mathop{\sim}\limits_{\scriptscriptstyle {#1\rightarrow\infty }}}}
\def\tozero#1{\mathrel{\mathop{\sim}\limits_{\scriptscriptstyle {#1\rightarrow0 }}}}
\def\slash#1{\mathord{\mathpalette\c@ncel#1}}
 \def\c@ncel#1#2{\ooalign{$\hfil#1\mkern1mu/\hfil$\crcr$#1#2$}}
\def\lsim{\mathrel{\mathpalette\@versim<}}
\def\gsim{\mathrel{\mathpalette\@versim>}}
 \def\@versim#1#2{\lower0.2ex\vbox{\baselineskip\z@skip\lineskip\z@skip
       \lineskiplimit\z@\ialign{$\m@th#1\hfil##$\crcr#2\crcr\sim\crcr}}}
\catcode`@=12 
\def\twiddles#1{\mathrel{\mathop{\sim}\limits_
                        {\scriptscriptstyle {#1\rightarrow \infty }}}}
\def\PR{{\it Phys.~Rev.~}}
\def\PRL{{\it Phys.~Rev.~Lett.~}}
\def\NP{{\it Nucl.~Phys.~}}
\def\NPBPS{{\it Nucl.~Phys.~B (Proc.~Suppl.)~}}
\def\PL{{\it Phys.~Lett.~}}
\def\PRep{{\it Phys.~Rep.~}}
\def\AP{{\it Ann.~Phys.~}}
\def\CMP{{\it Comm.~Math.~Phys.~}}
\def\JMP{{\it Jour.~Math.~Phys.~}}
\def\NC{{\it Nuov.~Cim.~}}
\def\SJNP{{\it Sov.~Jour.~Nucl.~Phys.~}}
\def\SPJETP{{\it Sov.~Phys.~J.E.T.P.~}}
\def\ZP{{\it Zeit.~Phys.~}}
\def\JP{{\it Jour.~Phys.~}}
\def\JHEP{{\it Jour.~High~Energy~Phys.~}}
\def\vol#1{{\bf #1}}\def\vyp#1#2#3{\vol{#1} (#2) #3}
\begin{titlepage}
\setcounter{page}{0}
\begin{flushright}
{\tt hep-ph/0104246}\\
{CERN/2001-114}\\
{Edinburgh 2001/02}\\
{RM3-TH 01/5}\\
\end{flushright}
\vskip.2cm
\begin{center}
{\Large \bf Extrapolating Structure Functions to Very Small $x$} \\

\vskip 0.8cm

Guido Altarelli,$^{1}$  Richard D. Ball$^{1,\;2}$  and
 Stefano Forte$^{3,\;}$\footnote[4]{On leave from INFN, Sezione di Torino, Italy}\\
\vskip.7cm
{\sl $^1$Theory Division, CERN\\ CH--1211 Geneva 23, Switzerland
\\ \vskip4pt{}$^2$Department of Physics and Astronomy, University of 
Edinburgh,\\ Mayfield Road, Edinburgh EH9
3JZ, Scotland\\
\vskip4pt {}$^3$INFN, Sezione di Roma Tre\\ Via della Vasca Navale
84, I-00146 Rome, Italy\\}
\end{center}

\vskip 1.5cm

\begin{abstract}

 We review small $x$ contributions to perturbative evolution
 equations for parton distributions, and their resummation. We
 emphasize in particular the resummation technique recently developed
 in order to deal with the apparent instability of naive small $x$ evolution
 kernels and understand the empirical sucess of fixed--order
 perturbation theory. We give predictions for the gluon distribution
 and the structure functions $F_2(x,Q^2)$ and  $F_L(x,Q^2)$ in an extended
 kinematic region, such as would be relevant for THERA or LEP+LHC $ep$ 
 colliders.

\end{abstract}

\vskip 1.2cm
\begin{center}
 to be published in
{\it the THERA book}
\end{center}

\end{titlepage}

Measurements of the inclusive structure functions $F_2(x,Q^2)$ and
$F_L(x,Q^2)$ at HERA have shown that the scaling violations of
structure functions are in extremely good agreement with the
perturbative next-to-leading order (NLO) QCD prediction, down to the
smallest values of $x$, and for all $Q^2\gsim 1\,\gev^2$~\cite{np:a666:129}. 
This agreement is surprising in that it is known  that perturbative
corrections beyond NLO in $\alpha_s$ are enhanced by powers of
$\xi\equiv\ln(1/x)$, and thus one would expect higher order
corrections to be sizable whenever $\alpha_s(Q^2) \xi\gsim 1$, i.e. in
most of the HERA kinematic region. Whereas techniques for the
inclusion of small $x$ contributions to leading twist evolution
equations have been known for some
time~\cite{pl:b351:313,pl:B348:582},
only recently did a consistent picture  of the general structure of these
contributions and their resummation emerge. Indeed, 
considerable theoretical progress has been spurred by the
determination~\cite{pl:b429:127}  of
next-to-leading corrections to the BFKL
kernel, which allows the computation of 
the next-to-leading $\log(1/x)$  (NLLx)
contributions to anomalous dimensions to all orders in
$\alpha_s$. Specifically, it is now
understood that the inclusion of NLLx
contributions leads to 
instability~\cite{pl:B341:161} of perturbative evolution, unless it is
suitably combined with a resummation of the collinear
singularities~\cite{jhep:07:019,np:B575:313,pr:D60:114036} which are
resummed
order by order in the standard QCD evolution
equations.  Furthermore, the NLLx perturbative corrections give rise
to increasingly large contributions to high orders of perturbation
theory~\cite{hep-ph-9805315,hep-ph-9806368} that make a nonsense of
the perturbative expansion and call for an all-order resummation of the
small-$x$ behaviour of the anomalous
dimensions~\cite{pl:b439:428,pl:B465:271}. 

Practical methods to deal with these issues have been developed
recently~\cite{np:B575:313,np:B599:383}, and lead to a resummation
prescription which is amenable
to numerical treatment and direct comparison with the data.
It then appears  that  the
observed smallness of perturbative higher order corrections at small
$x$ can be accommodated within the current knowledge of the general
structure of anomalous dimensions, but it poses very stringent constraints
on the form of the unknown higher order terms. Furthermore, even when
these constraints are respected, so that, as required by the data,
deviations of the behaviour of the observable structure
functions from the fixed
next-to-leading order prediction are very small, still 
non--negligible  modifications of the fitted parton distributions
at small $x$ are found. This, because of ambiguities in the
resummation procedure, entails larger uncertainties on parton
distributions at small $x$. Likewise, these corrections
have a sizable impact on the extraction of $\alpha_s$ from small $x$
data, both on the central value and the estimates of overall theoretical 
uncertainties~\cite{pl:b358:365,*hep-ph-9607289,np:B599:383}.

In the wider kinematic region available at THERA the small differences
between resummed and fixed--order predictions could be put to more
stringent tests. This would allow one to pin down more precisely the
ambiguities in the resummation procedure, thereby reducing the
uncertainty on parton distributions at small $x$ and on precision
determinations of $\alpha_s$ at small $x$. Also, the possibility of
reaching smaller values of $x$ for given $Q^2$ would allow a
test of resummed perturbation theory in a region where the relevant
resummation parameter $\alpha_s \xi$ is large, and also to see
whether the perturbative description of scaling violations remains
satisfactory or starts to break down, as is often  
suggested~\cite{hep-ph-9911289,*hep-ph-0102087}.

Here we briefly review our current understanding of
resummed perturbation theory at small $x$. We then give
predictions for the gluon distribution and the
structure functions $F_2$ and $F_L$ in two different
resummation scenarios, and compare these to fixed next--to--leading order
results in the kinematic range which is relevant for
THERA. This is essentially the same kinematic region accessible at 
a hypothetical lepton--hadron collider obtained combining LEP with 
the LHC, so our predictions would also be relevant at such a machine.

\section{Duality of small $x$ evolution}
\label{sec:abf:dual}
The basic result which allows the determination of contributions to
anomalous dimensions which are logarithmically
enhanced in $x$  to all orders in the
coupling, and thus their inclusion in evolution equations, is
the {\it duality} of perturbative
evolution~\cite{pl:b405:317,np:B575:313}: because leading--twist
evolution of structure
functions takes place  both in $x$ and $Q^2$, it admits a dual
description in terms of  equations for evolution in $t=\ln(Q^2/\mu^2)$ or evolution
in $\xi=\ln(1/x)$. 
This property
is easy to prove~\cite{pl:b405:317} when the coupling is fixed, and
can be shown to remain valid when the coupling runs by explicit
order--by--order perturbative computation~\cite{pl:B465:271}.

Let us first consider for simplicity the case (relevant in the very
 small $x$ limit) of a single parton distribution
$G(\xi,t)$, identified with the dominant eigenvector of perturbative
 evolution.  
The pair of dual evolution equations are then
\bea
\frac{d}{dt}G(\xi,t)&=&P(\xi,\as)\otimes G(\xi,t),
\label{eq:abf:tevol}\\
\frac{d}{d\xi}G(\xi,t)&=&K(t,\as)\otimes G(\xi,t),
\label{eq:abf:xevol}
\eea
The convolutions on the right--hand sides of the dual evolution equations 
(\ref{eq:abf:tevol}-\ref{eq:abf:xevol}) are with respect to $\xi$ 
in the first equation ($P(\xi,\as)$ is the usual splitting 
function) and with respect to
$t$ in the second equation; $\alpha_s=\alpha_s(t)$ and
is unaffected by convolutions.
Duality means that the solutions to these equations
coincide up to higher twist corrections provided the respective 
boundary conditions and kernels are suitably matched. 

The detailed form of the matching of boundary conditions
is irrelevant for our purposes, but it is important to
notice that the matching is such that
the boundary condition to~(\ref{eq:abf:tevol}) depends 
only on $\xi$ (and not on $t$)
and the boundary condition to~(\ref{eq:abf:xevol}) depends only on $t$
(and not on $\xi$) as required by factorization.
The matching of the kernels is given by the duality equation
\begin{equation}
\chi(\gamma(N,\as),\as)=N,\label{eq:abf:dual}
\end{equation}
or equivalently its inverse
\begin{equation}
\gamma(\chi(M,\as),\as)=M.\label{eq:abf:revdual}
\end{equation}
Here $\gamma$ is the usual anomalous dimension, related to the
splitting function by Mellin transformation with respect to $\xi$:
\begin{equation}
\gamma(N,\as)=\int^{\infty}_{0}\! d\xi\, e^{-N\xi}~P(\xi,\as).
\label{eq:abf:xmoms}
\end{equation}
The relation between $\chi(M,\as)$ and $K(t,\as)$ is somewhat more
complicated because, upon Mellin
transformation with respect to $t$
the running coupling $\as(t)$  on the right--hand side of
 Eq.~(\ref{eq:abf:tevol}) 
becomes a differential operator.
The relation between  the evolution kernel $K(t,\as)$ and the dual
kernel
$\chi(M,\as)$ can nevertheless be determined 
order by order in perturbation theory~\cite{pl:B465:271}: defining
\bea
K(t,\as)&=&\as K_0(t)+\as^2K_1(t)+\dots,\label{eq:abf:kerexp}\\
\chi(M,\as)&=&\as\chi_0(M)+\as^2\chi_1(M)+\dots,\label{eq:abf:chiexp}
\eea 
we get
\bea
&\chi_0(M)&=\int^{\infty}_{-\infty}\! dt\,
e^{-Mt}~K_0(t);
\nonumber \\
&\chi_1(M)&=\int^{\infty}_{-\infty}\! dt\, e^{-Mt}~K_1(t)+ 
{\beta_0 \over4\pi}  {1\over
2}{\chi_0(M)\chi_0''(M)\over{\chi_0'}^2(M)};\dots\>,
\label{eq:abf:effchi}
\eea
where $\beta_0=\smallfrac{11}{3}n_c-\smallfrac{2}{3}n_f$ is the
first coefficient of the QCD $\beta$ function. 

It follows from the form of the duality equation~(\ref{eq:abf:dual}) that
knowledge of the leading (next-to-leading,\dots) term in the expansion
of $\chi$ in powers of $\as$ at fixed $M$ determines the 
leading (next-to-leading,\dots)
term in the expansion of $\gamma$  in powers of $\as$ at fixed
$\as/N$: i.e. defining further
\bea
\gamma(N,\as)&=&\gamma_s(\smallfrac{\as}{N}) 
+\as\gamma_{ss}(\smallfrac{\as}{N})+\ldots,
\label{eq:abf:gamexp}
\eea
then
\begin{equation}
\chi_{0}(\gamma_{s}(\smallfrac{\as}{N}))={N\over\as},\quad
\gamma_{ss}(\smallfrac{\as}{N})
= -\frac{\chi_{1}(\gamma_{s}(\smallfrac{\as}{N}))}
{\chi'_{0}(\gamma_{s}(\smallfrac{\as}{N}))},\>\dots\>.
\label{eq:abf:dualexp}
\end{equation}
Likewise, knowledge of the leading,
next-to-leading,\dots terms in the expansion
of $\gamma$ in powers of $\as$ at fixed $N$ determines the 
leading, next-to-leading,\dots terms in the expansion of 
$\chi$  in powers of $\as$ at fixed $\as/M$: writing
\bea
\gamma(N,\as)&=&\as\gamma_0(N)+\as^2\gamma_1(N)+\dots,\>\\
\label{eq:abf:revgamexp}
\chi(M,\as)&=&\chi_s(\smallfrac{\as}{M})
+\as\chi_{ss}
(\smallfrac{\as}{M})+\ldots,
\label{eq:abf:revchiexp}
\eea
then
\begin{equation}
\gamma_{0}(\chi_{s}(\smallfrac{\as}{M}))={M\over\as},\quad
\chi_{ss}
(\smallfrac{\as}{M})= -\frac{\gamma_{1}(\chi_{s}(\smallfrac{\as}{M}))}
{\gamma'_{0}(\chi_{s}(\smallfrac{\as}{M}))},\>\dots\>.\label{eq:abf:revdualexp}
\end{equation}
It should be understood that the running coupling corrections 
Eq.~(\ref{eq:abf:effchi}) are always included in the definition of 
$\chi$ in the above equations.

Because the $\xi$ evolution equation is essentially the same as the
BFKL equation (up to factorization scheme and scale choices, which
become relevant beyond leading order~\cite{pl:b429:127,pl:b430:349})
the duality relation can be viewed as a consistency
condition between this equation and the standard renormalization 
group equation for
moments of structure functions in the region of their common validity
(i.e. large $Q^2$ and small $x$). Hence, knowledge of
the BFKL kernel $K(t,\as)$~(\ref{eq:abf:kerexp}) can be translated into information of
$\chi$~(\ref{eq:abf:chiexp}),  which in turn can be used to gain
information on the logarithmically enhanced contributions $\gamma_s$,
$\gamma_{ss}$,\dots~(\ref{eq:abf:gamexp}) to the anomalous dimension $\gamma(\as,N)$, and
conversely.
In fact, 
the leading-order equation in~(\ref{eq:abf:dualexp}) has
been known for a long time~\cite{pl:B116:291}; the new insight here is
that this is just a consequence of a more general duality.
\section{The Double--Leading expansion}
\label{sec:abf:dlead}

\begin{figure}[t!]
\begin{center}
\epsfig{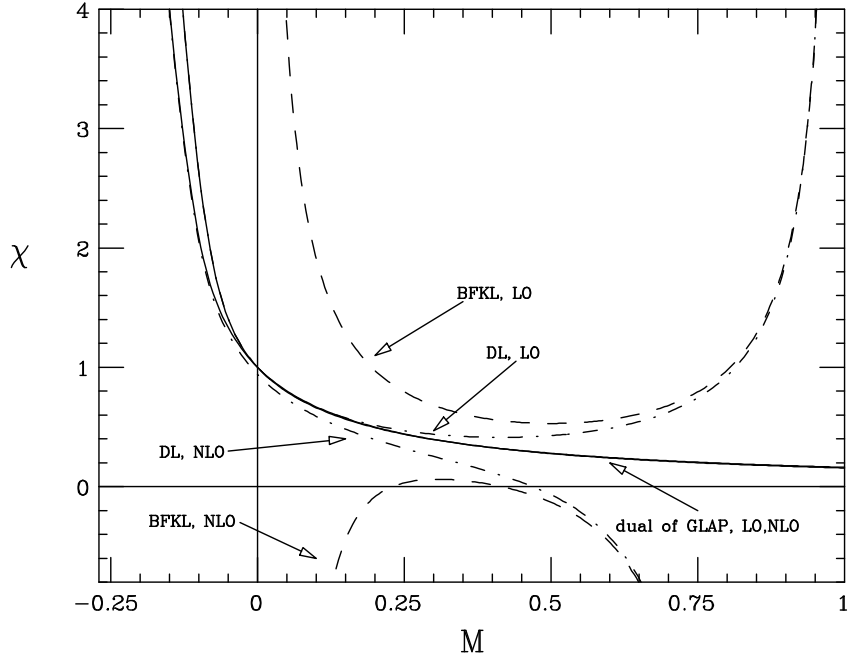}
\end{center}
\caption{Plots of different approximations to 
$\chi$: the BFKL leading and
next-to-leading  order functions~(\ref{eq:abf:chiexp}), $\as\chi_0$
and $\as\chi_0+\as^2\chi_1$ (dashed); the LO and NLO dual $\as
\chi_{s}$  and $\as \chi_{s}+ \as^2
\chi_{ss}$~(\ref{eq:abf:revchiexp})   of the one and two
loop anomalous dimensions (solid), and the double--leading
functions at LO and NLO defined in  Eq.~(\ref{eq:abf:cdl}) (dotdashed).
All curves are computed with $\alpha_s=0.2$.}
           \label{fig:abf:chi}
\end{figure}
Only the first two orders in the expansion of $\chi$ at fixed $M$ and
$\gamma$ at fixed  $N$ are currently known. While the perturbative
expansion of $\gamma$ is well-behaved, in the sense that $\as\gamma_1$ is
a small correction to $\gamma_0$ for reasonable values of the coupling
constant, the perturbative expansion of $\chi$ is very poorly behaved,
in that the NLO correction $\chi_1$ 
completely changes the qualitative shape of
the kernel. In particular (see Fig.~\ref{fig:abf:chi}), 
in the physical region $0\le M\le1$  the LO
kernel has simple poles with positive residue at $M=0$ and $M=1$ and a
minimum in between. The NLO correction $\chi_1$ instead has higher order
poles with negative coefficient, and, for 
 any realistic value of $\as$ (essentially, for all $\as\gsim 0.03$)
the full NLO function 
 has just a 
maximum (for smaller $\alpha_s$ it has a minimum and two 
maxima)~\cite{hep-ph-9805315}. It
is easy to show that the solution to the evolution equation determined
by a kernel with this shape displays unphysical
oscillatory behaviour
in the limit as $x\to0$, and thus, in particular,  leads to negative
cross-sections~\cite{pl:B341:161}.

Because the
Mellin transform~(\ref{eq:abf:effchi}) of $t^k=\ln^k(Q^2/\mu^2)$ is
$(k-1)!/M^{k-1}$, the presence of $1/M$ poles in the
kernel $\chi$ is related to collinear singularities: indeed, according
to Eq.~(\ref{eq:abf:revdual})
the coefficients
of these singularities are determined by knowledge of the 
anomalous dimensions $\gamma_0$, $\gamma_1$,\dots 
in the usual renormalization group equations, which
resum collinear singularities. It is easy to
understand~\cite{np:B575:313}  why these
singularities lead to a series of poles in $M=0$ with alternating
signs. Indeed, recall that momentum conservation implies that the
largest eigenvalue of the anomalous dimension matrix vanishes at
$N=1$, i.e. $\gamma(1,\as)=0$, which by duality~(\ref{eq:abf:dual})
implies $\chi(0,\as)=1$. It follows that if, in the vicinity of $M=0$,
$\chi_s$ behaves as 
\begin{equation}
\chi_s\tozero M {\as\over\as+M}={\as\over M}-{\as^2\over M^2}+{\as^3\over
M^3}+\dots\quad:
\label{eq:abf:momcons}
\end{equation} 
the series of poles in $\chi_s$,
$\chi_{ss}$,~\dots actually sums up to the regular behaviour 
$\chi(0,\as)=1$.

The poles in $\chi$  as $M\to 0$ are summed to all orders into
$\chi_s$, $\chi_{ss}$, \dots, and thus the undesirable behaviour of the
expansion of $\chi$ can be removed by defining
order by order an improved expansion. Namely,
we define a {\it double leading expansion} where  to
each order in $\as$  both the terms present in the expansion in powers 
of $\alpha_s$  at
fixed $M$  and at fixed $\as/M$ are included:
\bea &\chi(M,\as)&=\left[\as\chi_{0}(M) +\chi_{s}\left({\smallfrac{\as}{M}}\right)-
\smallfrac{n_c\as}{\pi M}\right]\nonumber\\ &&\qquad +\as\left[\as\chi_{1}(M)
+\chi_{ss}\left({\smallfrac{\as}{M}}\right)-\as\left(\smallfrac{f_2}{M^2}+
\smallfrac{f_1}{M}\right)-f_0\right]+\cdots\>.
\label{eq:abf:cdl}
\eea 
In this expansion, in the vicinity of $M=0$ the singularities of
$\chi_0$, $\chi_1$,\dots are resummed into $\chi_s$, $\chi_{ss}$, while 
the subtraction terms  avoid double--counting of these contributions.
Note (see Fig.~\ref{fig:abf:chi}) that at larger  values
of $M$ the shape of  $\chi_0$, $\chi_1$, \dots is reproduced, but in
most of the $M$ range the kernel Eq.~(\ref{eq:abf:cdl}) coincides
with the (dual of) the standard anomalous dimensions $\gamma_0$ and
$\gamma_1$, consistent with the empirical smallness of small-$x$
correction to perturbative evolution.
This also has the significant implication that the double--leading
expansion of $\chi$ is as stable as the usual expansion of $\gamma$ at
fixed $N$. 

It is easy to show that the 
corresponding double leading expansion of $\gamma$,
\bea &\gamma(N,\as)&=\left[\as\gamma_{0}(N) +\gamma_{s}\left(\smallfrac{\as}{N}\right)-
\smallfrac{n_c\as}{\pi N}\right]\nonumber\\ &&\qquad +\as\left[\as\gamma_{1}(N)
+\gamma_{ss}\left(\smallfrac{\as}{N}\right) -\as\left(\smallfrac{e_2}{N^2}+
\smallfrac{e_1}{N}\right)-e_0\right]+\cdots,\label{eq:abf:gdl}
\eea 
is consistent with duality,
in that $\chi$~(\ref{eq:abf:cdl}) and $\gamma$~(\ref{eq:abf:gdl}) are 
dual to each other order by order in the double--leading expansion, 
up to higher order corrections. Hence, for practical applications we  may 
directly use the double--leading anomalous dimension~(\ref{eq:abf:gdl}) in
the usual evolution equation~(\ref{eq:abf:tevol}). This will ensure that
collinear singularities are resummed according to the renormalization
group in the usual way, while leading logs of $1/x$ are consistently
included up to next--to--leading order.

For actual phenomenology, the full set of anomalous dimensions  and
coefficient functions are needed. It is easy to see that the
double--leading expansion is consistent with diagonalization of the
anomalous dimension matrix, in the sense that
one may equivalently, up to subleading corrections, 
construct a two by two matrix of double--leading 
anomalous dimensions and diagonalize it, or else construct directly a
double--leading expansion of 
eigenvalues and projectors. Because one of the two eigenvectors of
$\gamma$ is
free of small--$x$ singularities, so its double--leading expansion
coincides with the standard expansion at fixed $N$, the latter
procedure is in practice simpler. Hence, the double leading expansion
can be fully defined in terms of the expansion of the large anomalous
dimension eigenvalue, and of the quark--sector matrix elements which
determine the projectors on the eigenvectors. Likewise, one can construct
double--leading coefficient functions, and prove that the expansion
transforms consistently upon changes of factorization scheme.
Detailed proofs and results needed for a practical implementation are
given in ref.~\cite{np:B599:383}.

\section{Resummation}
\label{sec:abf:resumm}

Even though the difference between the  double--leading expansions of 
$\chi$~(\ref{eq:abf:cdl}) and
$\gamma$~(\ref{eq:abf:gdl}) is subleading, it can in
practice be large when $M\gsim 0.25$. Indeed, recall that
duality~(\ref{eq:abf:dual}) implies $\chi=N$. It is clear from 
Fig.~\ref{fig:abf:chi} that in the region $M\gsim 0.25$
the difference between
the leading order and next-to-leading order
 double--leading curves is small for any fixed
value of $M$, but it is quite large for a fixed value of $\chi=N$, 
because  the  curves are 
almost parallel to the $M$--axis: the LO BFKL curve has a minimum at 
$M=1/2$. Since $\gamma$ is a function of $N$,  in this region the
perturbative solution~(\ref{eq:abf:dualexp}) of the duality
relation~(\ref{eq:abf:dual}) is not good and the expansion of $\gamma$
is not well behaved. 

A possible way out is to determine the double--leading
$\gamma$~(\ref{eq:abf:gdl}) 
from the double--leading $\chi$~(\ref{eq:abf:cdl}) 
by solving the duality relation~(\ref{eq:abf:dual}) exactly (rather
than perturbatively). This can be done for instance by numerical
methods, or equivalently by
differentiating with respect to $t$  
the solution of the evolution
equation~(\ref{eq:abf:xevol}) determined using
the double--leading $\chi$
kernel~(\ref{eq:abf:cdl}), as in ref.~\cite{pr:D60:114036}.
However, this approach, besides being cumbersome to implement in
standard evolution codes, has the shortcoming that it hides a genuine
perturbative ambiguity. Indeed, in this way the
perturbative expansion of $\gamma$ is in practice
stabilized by assuming that
in the region $M\approx1/2$ the (large)
subleading corrections to $\gamma$ will be such as to reproduce the 
shape of $\chi$, as computed to some fixed perturbative order, or
possibly further improved according to a model of its behaviour at
large $M\sim 1$~\cite{pr:D60:114036}. 

We instead prefer to use only the available perturbative
information on $\gamma$, without making model--dependent
assumptions. It can be shown~\cite{pl:B465:271}
that the poor perturbative behavior of the expansion of $\gamma$
at fixed $\as/N$ manifests itself in a rise of the associate
splitting functions: $P_{ss}/P_s\toinf\xi \as \xi$, 
$P_{sss}/P_s\toinf\xi \as^2 \xi^2$ and so on. 
This rise can be removed by simply subtracting at each
order a suitable
constant $c_i$ from $\chi_i$ (computable order by order in perturbation
theory as a function of $\chi_i$ and their derivatives at $M=1/2$), 
and then determining
$\gamma_{ss\dots}$ from the subtracted $\chi_i$. Thus, the expansion
of $\gamma$~(\ref{eq:abf:chiexp}) can be stabilized by just
 reorganizing  the perturbative expansion of $\chi$:
\bea
\chi(M,\as)&=&\as \chi_0(M)+\as^2\chi_1(M)+\dots\label{eq:abf:curexp}\\
&=&\as \tilde \chi_0(M) +\as^2\tilde\chi_1(M)+\dots,
\label{eq:abf:clamsub}
\eea
where
\begin{equation}
\as\tilde \chi_0(M,\as)\equiv \alpha_s
\chi_0(M)+\Delta \lambda,\qquad 
\tilde\chi_i(M)\equiv\chi_i(M)-c_i,
\label{eq:abf:ctil}
\end{equation}
for $i=1,2,\dots$, and thus
\begin{equation}
\Delta\lambda\equiv \sum_{n=1}^\infty \alpha_s^{n+1} c_n.
\label{eq:abf:dellam}
\end{equation}
If $\chi$ has a minimum, then its value  at the
minimum coincides~\cite{np:B575:313}  with the value of 
$\tilde\chi_0$ at its minimum $M=1/2$, namely
\begin{equation}
\lambda \equiv\tilde\chi_0(\half)=\chi_0(\half)+\Delta \lambda.
\label{eq:abf:lamdef}
\end{equation}
Since the
value of $\chi$ at its minimum determines the asymptotic behaviour of
the structure function as $x\to0$, 
this implies that in order to remove the perturbative instability it is
necessary and sufficient to resum the asymptotic small $x$ behaviour into the
leading order kernel $\tilde\chi_0$. 
The perturbative 
instability signals the fact that the all--order asymptotic 
behaviour must be known to all orders.

Of course, we are free to use any particular truncation of
$\Delta \lambda$~(\ref{eq:abf:dellam}): for instance, we could
simply take $\chi$ to coincide with its NLO form in the
double--leading expansion. Eq.~(\ref{eq:abf:clamsub}) then provides
us with a stable perturbative expansion of $\gamma$, which at NLO is
very close to the exact dual of $\chi$, the large subleading
corrections having been resummed in a minimal way. In this way
Eq.~(\ref{eq:abf:clamsub}) gives us a simple prescription
which completely stabilizes the double--leading expansion of $\gamma$
whenever the double--leading expansion of $\chi$ is also
stable. Hence, any specific resummation of $\chi$ (such as that
constructed in ref.~\cite{pr:D60:114036}) can be accommodated 
in this formalism. Since 
however we prefer not to rely on such specific assumptions, we will 
consider $\lambda$~(\ref{eq:abf:lamdef})
as a free parameter.

To NLO, the (resummed) expansion of $\gamma$ obtained from
Eq.~(\ref{eq:abf:clamsub})  is
related to the unresummed expansion obtained from Eq.~(\ref{eq:abf:curexp})
by
\begin{equation}
\tilde\gamma(N,\as)=\tilde\gamma_s\left(\smallfrac{\as}{N}\right)+
\as\tilde\gamma_{ss}\left(\smallfrac{\as}{N}\right)
+\dots,\label{eq:abf:gamimpr}  
\end{equation}
where 
\bea
\tilde\gamma_s\left(\smallfrac{\as}{N}\right)
&=&\gamma_s\left(\smallfrac{\as}{N-\Delta \lambda}\right),\nonumber\\
\tilde\gamma_{ss}\left(\smallfrac{\as}{N}\right)&=&
 \gamma_{ss}\left(\smallfrac{\as}{N-\Delta \lambda}\right) 
-{\chi_1(\half)\over\chi'_0\left(\gamma_s
\left(\smallfrac{\as}{N-\Delta \lambda}\right)\right)}.
\eea
Since the resummation only involves formally subleading terms, 
\begin{equation}
\gamma_s+\as\gamma_{ss}=\tilde\gamma_s+\as\tilde\gamma_{ss}+O(\as^3/N).
\label{eq:abf:expequiv}
\end{equation}

A resummed double--leading expansion can finally be constructed by
combining the resummed anomalous dimension $\tilde
\gamma$~(\ref{eq:abf:gamimpr}) with the standard expansion of $\gamma$
at fixed $N$. This gives a resummed double--leading expression for
the large anomalous dimension eigenvector. It can further be shown~\cite{np:B599:383} that
resummed double--leading expressions for the full matrix of anomalous
dimensions and for coefficient functions can be obtained by performing
the replacement $N\to N-\Delta\lambda$ in all remaining quantities,
i.e. the projectors and the coefficient functions. 

The construction of the resummed double--leading expansion entails a
further ambiguity in the treatment of the double counting
subtractions in Eq.~(\ref{eq:abf:gdl}):
because these terms are common to the fixed--$N$ expansion
$\gamma_0$, $\gamma_1$,\dots and the fixed $\as/N$ expansion
$\gamma_s$, $\gamma_{ss}$,\dots,  we are free to
decide whether to leave them unaffected by the replacement
$N\to N-\Delta\lambda$ or not. This defines a pair of resummation
procedures, which of course only differ by subleading terms. Clearly,
a variety of intermediate alternatives would also be possible. The
main difference between these prescriptions is the nature of the small
$N$ singularities of the anomalous dimension, which control the
asymptotic small $x$ behaviour.
The resummed anomalous dimension 
always has a cut starting
at $N=\lambda$ Eq.~(\ref{eq:abf:lamdef}), which
corresponds~\cite{np:B599:383}  to an $x^{-\lambda}$  behaviour
of splitting functions at small $x$. If the subtractions are affected by
the replacement, then $\gamma_0$ and $\gamma_1$ are the same as in the
unresummed case (S--resummation), i.e. they
have  a simple pole at $N=0$, which leads to a
``double--scaling''\cite{pr:10:1649,*pl:b335:77} 
rise at small $x$. If the subtractions are unaffected, this pole is
removed by the subtraction itself (R--resummation). 
It follows that if $\lambda$ is positive, then the two
resummations give similar results at small $x$, namely an $x^{-\lambda}$ 
power rise. If
$\lambda\leq 0$, the
S--resummation will display  double scaling at small $x$, while
the  R--resummation will display
a valence-like $x^{-\lambda}$ behaviour.

\section{Predictions for THERA}
\label{sec:abf:pred}

A comparison of the resummation discussed in the
previous sections to recent HERA data~\cite{hep-ex-0012053} was
presented in ref.~\cite{np:B599:383}. The best--fit results of that
reference can be used to obtain predictions for THERA and discuss the
 study of small $x$ scaling violations at such a facility. Because of the
larger center--of--mass energy available at THERA, these 
predictions essentially amount to an extension of the
current kinematic range of $1/x$ by about a decade for each value of $Q^2$. It
is interesting to note that more or less the same center--of--mass
energy would be available at a hypothetical lepton--hadron collider
obtained combining LEP with the LHC.

The phenomenological analysis of ref.~\cite{np:B599:383} is based on a
fit to data for the reduced cross--section
\begin{equation}
\sigma_{red}(x,y,Q^2)=F_2(x,Q^2)+{ y^2\over 2(1-y)+y^2}
F_L(x,Q^2).
\label{eq:abf:sigred}
\end{equation}
determined from structure functions $F_2$ and $F_L$ 
 computed to next--to--leading
order in the double leading expansion with the R-- and S--resummation
prescriptions discussed in Sect.~\ref{sec:abf:resumm}, by evolving
parton distributions given at a  scale $Q_0=2$~GeV. A standard
unresummed next--to--leading order 
fit is also performed for comparison. The fits are performed in the
parton scheme, so the quark distribution coincides by construction
with $F_2$. The large--$x$
shape of parton distributions is taken from a global fit, while the
small $x$ behaviour is parametrized by two free parameters $\lambda_q$
and $\lambda_g$, which give the asymptotic small--$x$ behaviour of the
singlet quark and gluon distributions respectively as
$x^{-\lambda_q}$, $x^{-\lambda_g}$. The resummation parameter $\lambda$
Eq.~(\ref{eq:abf:lamdef}) is also left as a free parameter. The strong
coupling is fixed at $\as(M_z)=0.119$. 
\begin{figure}[t!]
 \begin{center}
 \epsfig{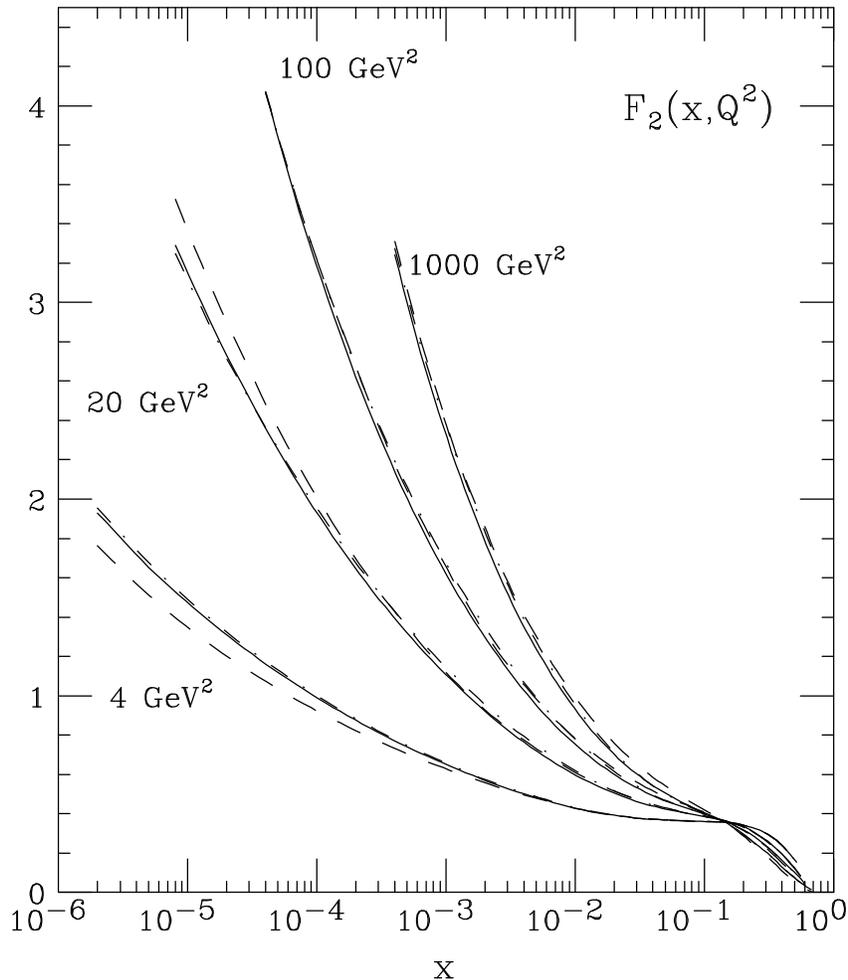}
 \end{center}
 \caption{The structure function $F_2(x,Q^2)$ obtained from a
 fit~\protect\cite{np:B599:383} to HERA
 data~\protect\cite{hep-ex-0012053}. The prediction for THERA is the   
last decade in $x$
 for each value of $Q^2$. The solid curve is an unresummed 
fixed--order two loop fit, while the dot-dashed curve corresponds to  
 the S--resummation
 and the dashed curve to the R--resummation discussed in
 Sect.~\ref{sec:abf:resumm}. }
 \label{fig:abf:ftwo}
\end{figure}

Because at the initial scale
$Q_0$ abundant data are available down to the smallest values of $x$,
and $F_2$ coincides with the quark distribution, 
the quark exponent $\lambda_q$ turns out to be the same
in all fits, and gives the effective power rise of the
$F_2(x,Q^2_0)\tozero{x} x^{-\lambda_q}$: $\lambda_q\approx 0.2$.
The best--fit value of the gluon exponent is valence-like in all
fits: in the two-loop fit it is $\lambda_g\approx-0.1$; while in
the resummed fits it is significantly more valence-like, $\lambda_g\approx-0.2$
for both S-- and R--resummation. The value of the resummation
parameter $\lambda$ instead varies significantly
according to the resummation
prescription which is adopted. 
For the S--resummation, any value $\lambda \leq 0$ gives
a good fit, with the best fit around $\lambda\approx -0.25$. As
discussed in Sect.~\ref{sec:abf:resumm} the S--resummation with
vanishing or
negative $\lambda$ is closest to the unresummed fixed--order result.
With the R--resummation, instead, only a fine--tuned value of
$\lambda\approx0.2$ gives a good fit. This  value of
$\lambda$ turns out to be the same which one gets by  fine--tuning
the resummed anomalous dimension so that it be closest to the unresummed
one in the HERA kinematic region~\cite{np:B575:313}. Both resummed
fits give a similar $\chi^2\approx 52$ with $93$ degrees of freedom,
to be compared to the unresummed value $\chi^2=60$.
\begin{figure}[t!]
 \begin{center}
 \epsfig{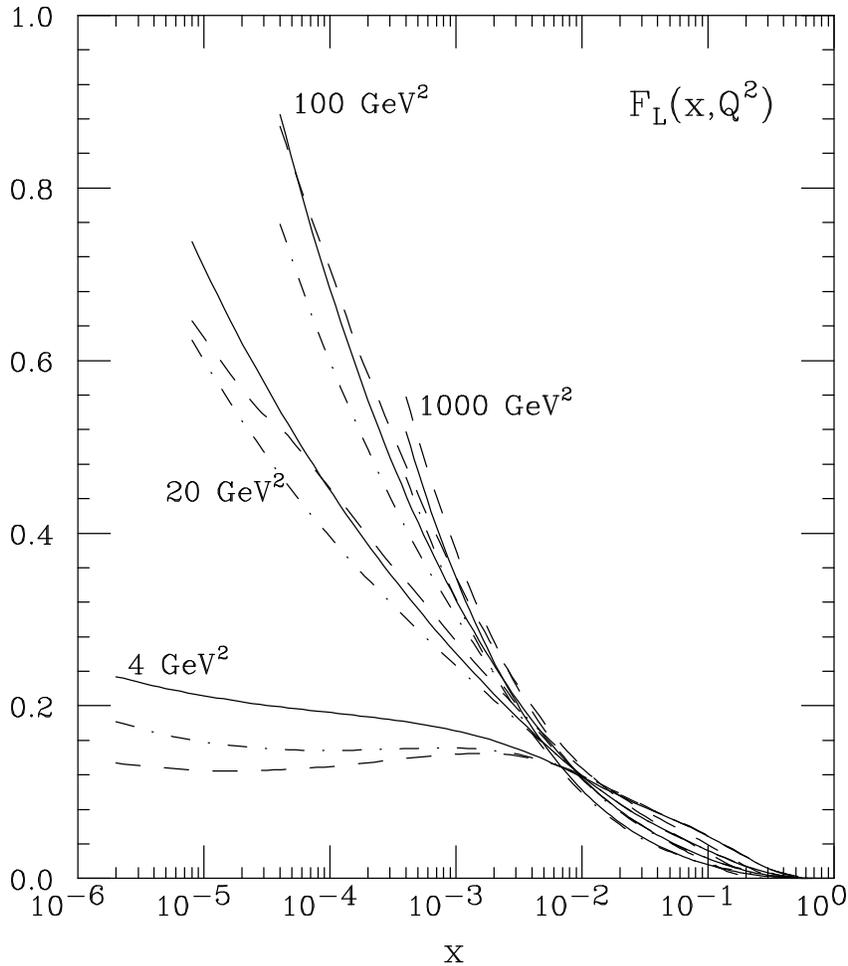}
 \end{center}
 \caption{Same as Fig.~\ref{fig:abf:ftwo}, but for
 the structure function $F_L(x,Q^2)$.}
 \label{fig:abf:fl}
\end{figure}

The structure function $F_2(x,Q^2)$ obtained in these fits is
displayed in Fig.~\ref{fig:abf:ftwo}. It is apparent that, given the
high precision of the HERA data, all curves, which give good fits to
the data, are constrained to lie essentially on top of each other
throughout the HERA region, except possibly at the smallest $x$
values $x\lsim10^{-4}$ at the initial scale $Q_0$, where the
R--resummation curve rises slightly less. In the THERA range it is
still very difficult to tell the difference between various
prescriptions at higher scales $Q^2\ge100$~GeV$^2$, but at lower
scales, while results
in the S--resummation are still essentially indistinguishable from the
two--loop ones, the R-resummation predicts a somewhat faster evolution.
\begin{figure}[t!]
 \begin{center}
 \epsfig{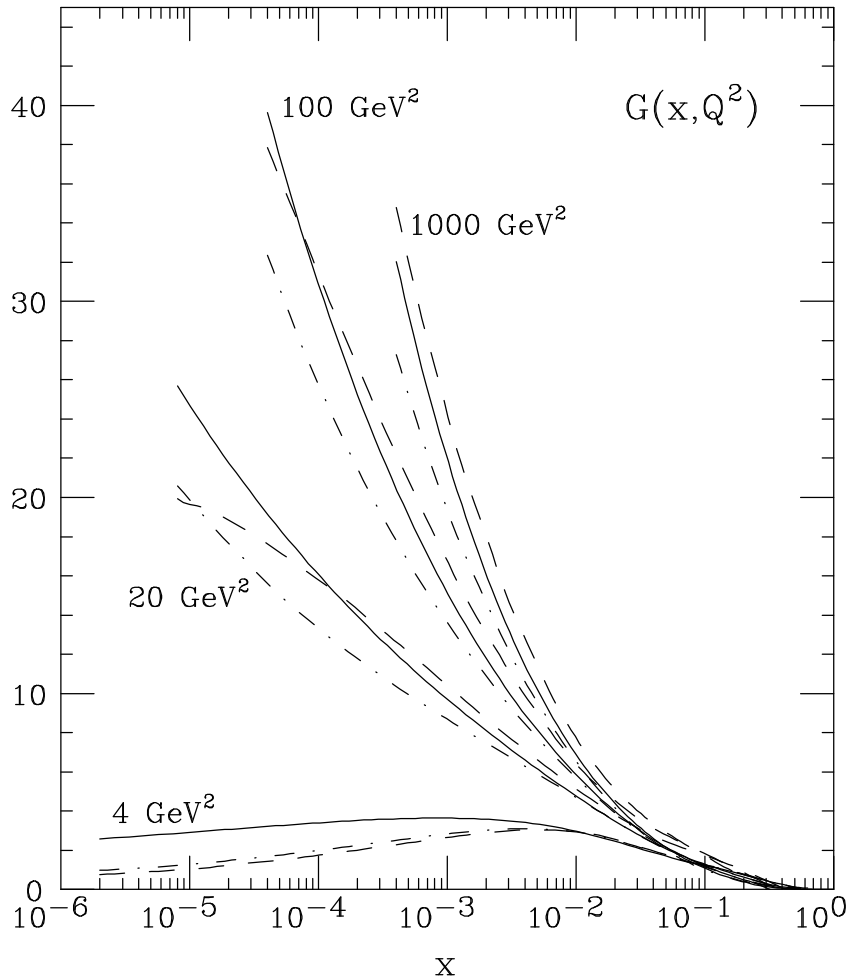}
 \end{center}
 \caption{Same as Fig.~\ref{fig:abf:ftwo}, but for
 the \MS\ gluon distribution $G(x,Q^2)=x g(x,Q^2)$.}
 \label{fig:abf:gluon}
 \end{figure}

The structure function $F_L$ is displayed in
Fig.~\ref{fig:abf:fl}. This structure function is not determined very
accurately by the HERA data for the reduced cross section
Eq.~(\ref{eq:abf:sigred}),  essentially because of the scarcity
of large--$y$ data. The spread of the results is accordingly
larger. Because $F_L$ at small $x$ has a large gluonic component, the
behaviour of $F_L$ is 
similar to that of the gluon distribution, displayed in
Fig.~\ref{fig:abf:gluon}. For ease of comparison with other work, the
\MS\ gluon is shown, even though our fits were perfomed in the DIS
scheme. 
Both resummations give a
rather softer behaviour than the fixed--order one
at the initial scale $Q_0$: valence-like for
the gluon, turned into a rise of $F_L$ at very small $x$ (well into
the THERA range) by the rise 
of the coefficient function. 
The R-resummation, however, then leads to
significantly more rapid evolution: as the scale increases, the
resummed gluon overtakes the fixed--order one. This is essentially due
to the fact that the R--resummation eventually generates an $x^{-\lambda}$
power behaviour of all parton distributions at small $x$, 
and here $\lambda=0.2$ (power rise). The 
S--resummation, instead, leads to evolution dominated by double
scaling, 
which is very similar to
the fixed--order one, and thus both the gluon and $F_L$ preserve the
relative softness that they displayed at the initial scale.

Summarizing, it is clear that resummation effects, though very small,
become increasingly important as $x$ decreases. Deviations from the
fixed--order behaviour appear, at least in a simultaneous
determination of $F_2$ and $F_L$. At present, the
deviations from the fixed order prediction are within the
uncertainties of the resummation procedure: so, while it is clear that
the resummed gluon distribution is softer than the unresummed one, it
is hard to tell whether it will evolve faster or slower at small $x$. 
The underlying physics between these options is quite different:
either the onset of a slow power-like rise (R-resummation), or
persistence of the double--scaling rise (S--resummation). 
Both possibilities are consistent with present-day data, as well as with
our current knowledge of anomalous dimensions. Understanding which (if
any) of these possibilities is correct could be of considerable
theoretical interest, and in particular, it could shed light on  the
running of the
coupling in the high--energy limit~\cite{pl:b405:317,np:B599:383}.

In conclusion, accurate data in the THERA 
region could reveal significant
differences between the resummations procedures, and thus shed light on
the structure of unknown higher order contributions to perturbative
anomalous
dimensions, and on the underlying physics. The simultaneous
measurement of $F_2$ and $F_L$ in a wide range of $Q^2$ at small $x$ 
would allow an accurate determination of structure functions at small
$x$ which are required e.g. for precise phenomenology of heavy quark
production at future colliders. 

\bigskip\noindent
{\bf Acknowledgements}: This work was supported in part by
EU TMR  contract FMRX-CT98-0194 (DG 12 - MIHT).
\bigskip

{\raggedright
\bibliography{tprep}
}
\end{document}